# Towards data-driven modeling and real-time prediction of solar flares and coronal mass ejections

M. Rempel, Y. Fan, M. Dikpati, A. Malanushenko (HAO/NCAR),
M. D. Kazachenko (CU/NSO), M. C. M. Cheung, G. Chintzoglou (LMSAL),
X. Sun (U. of Hawaii), G. H. Fisher (U. of Berkeley), T. Y. Chen (Columbia)

**Modeling of flare and CME events:**

Modeling of transient events in the solar atmosphere requires the confluence of 3 critical elements: (1) Models with a sufficient sophistication in terms of physics and their ability to simulate a solar-like setup in terms of domain extent and time-scales; (2) The availability of data (combination of remote sensing with in-situ observations) with a stable quality, time-duration, spatial coverage and cadence; (3) The ability to ingest these data into models and to continuously update and correct the model state to reflect the observed conditions on the Sun.

**Current state:**

Solar eruptions such as flares and coronal mass ejections (CMEs) are manifestations of the explosive release of magnetic energy stored in the current carrying (twisted) coronal magnetic fields. Determining the evolution of the coronal magnetic field is therefore critical for enabling the prediction of time of eruptions and their geo-effectiveness. There are primarily two groups of models that simulate evolving coronal magnetic fields in 3D: quasi-static (or *time-independent*) and dynamic (*time-dependent*). In the quasi-static group, the potential, linear, and non-linear force-free field (NLFFF) extrapolations have been developed. These models apply the vacuum-limit assumption, which reasonably assumes that magnetic pressure dominates the gas pressure (low-$\beta$ regime). Another group of models, that use fewer approximations, are time-dependent models. The time-dependent models come in a variety of setups. Smaller-domain but high-resolution models, such as Bifrost (e.g. Gudiksen et al. 2011) and MURaM (e.g. Rempel 2017, Cheung et al. 2019), have the most comprehensive physics and allow for forward modeling of observables from visible to EUV and soft X-ray wavelengths, but are computationally expensive. Large-scale models, including global models, have lower resolution and simplified physics, but allow for the modeling of the evolution of individual CME events in the global solar corona out to large radial distances from the Sun. The ingestion of observations is critical for all models, in order to allow for the modeling of actual solar events. The research focus is currently moving rapidly from the domain of data-inspired models (where the initial setup mimics certain properties observed on the Sun) to data-driven models that rely on routine vector magnetic field observations from the HMI/SDO and require new approaches to derive electric fields (or plasma velocities) from these observations.

Currently, magnetohydrodynamic (MHD) models of CME events typically derive *boundary conditions* for the magnetic field from the observed photospheric magnetograms and produce the *pre-eruptive configuration* using (1) some form of boundary driving of magnetic flux, such as flux emergence, shear flows, and helicity condensation (e.g. Cheung and DeRosa 2012, Jiang et al. 2016, Fan 2016, Mackay et al. 2018), (2) nonlinear force-free field (NLFFF) extrapolations (e.g. Guo et al. 2019), or (3) analytical flux-rope models that are inserted into the source region of the eruption (e.g. Toeroek et al. 2018). In many cases, the lower boundary driving and/or the construction of the force-free field and the inserted flux ropes are still largely *ad hoc* and not well constrained by observations. As a result, generally only qualitative agreement is obtained between the modeled magnetic field evolution and the observed event. Truly quantitative models of eruptive flare and CME events that are well constrained by observations are yet to be developed.

Another challenge is bridging the gap between physically required and numerically feasible resolution and cadence. Lately, two types of data-driven models have been developed for this purpose: magneto-frictional (MF, e.g. Cheung and DeRosa 2012) and MHD models (e.g. Hayashi et al. 2018). The MF model assumes that the plasma velocity in the induction equation is proportional to the local Lorentz force; the subsequent plasma evolution leads to a relaxation of a magnetic configuration toward a force-free state. MF is more computationally efficient than MHD and is suitable for description of the slow quiescent evolution of active regions (ARs), but not for modeling of the flares. The MHD models explicitly solve a full set of MHD equations including the plasma properties. The MHD approach is suitable for modeling the rapid evolution of ARs during flares but is too computationally expensive to model their long-term quiescent evolution. A hybrid framework, where the MF model is used to model quiescent periods of AR evolution and the MHD model is used to model flaring periods of AR evolution, has been recently developed within the Coronal Global Evolutionary model (CGEM, Hoeksema et al. 2020). Other global approaches use locally concentrated or adaptively refined grids (e.g. the SWMF, Tóth et al. 2005). The future lies in data-driven models that can be implemented through: (1) boundary driving, the use of temporal sequence of photospheric electric fields (derived from vector magnetograms to represent the realistic flux transport) at the lower boundary for a time-dependent coronal field model (e.g. Fisher et al. 2015); and (2) data assimilation, the use of temporal sequence of observations for updating the physical state of a model through statistical methods such as Ensemble-Kalman filters (EnKF).

*Data Assimilation* (DA) is widely used and well established in the Earth atmospheric community; however, the use in solar physics is currently limited to applications of solar-cycle forecasting (e.g. Dikpati et al. 2016, Kitiashvili 2016). The full implementation of DA (through EnKF) in MHD models is a step beyond boundary driving which provides the following advantages: (1) evolution of an ensemble model that can account for observation uncertainties and calculate model errors; (2) correction of the full model state in response to new observations; (3) the ability to assimilate a wide variety of observations, including remote sensing and in-situ observations, not limited to just the lower boundary of the system. However, EnKF DA requires substantial computing cost due to the need of (1) adequate ensemble runs and (2) computation of observables from physical model-outputs to compare the model with real observations at every assimilation-step.

**Future developments and enabled science:**

Over the coming decades the critical progress will be enabled by: (1) Improving models such that also large-scale simulations of the solar atmosphere allow for a **detailed modeling of processes and the forward modeling of observables** through a combination of implemented physics and numerical resolution; (2) The **full implementation of DA** to ingest a wide range of remote sensing and in-situ observations from heliospheric observatories; (3) The use of **accelerator technologies such as GPUs** to boost the computation speed. This will allow models to run faster than real time in order to enable research on a large number of observed events and allow for operational space weather modeling. These developments are only possible if the field widely adopts the latest computing technologies (such as the use of accelerators in form of GPUs) and stays on the forefront of new developments. While some models have been refactored for GPU use (e.g. MAS, Caplan et al. 2019) or are in the process of refactoring (e.g. MURaM), the field of solar physics overall is behind the curve in adopting GPU computing; (4) The recent progress in data-driven models has relied critically on **routine vector magnetic field observations** from the HMI/SDO and new approaches to derive electric fields (or plasma velocities) from these observations. The need for such observations will only increase once full DA approaches are incorporated into MHD models; uncertainties, in particular systematics in current and future observations, will need to be better quantified. Future observations of **vector magnetic fields in the chromosphere and transition region** can provide the much needed, more NLFFF-consistent lower boundary conditions for both the force-free-field construction of the pre-eruption coronal magnetic field and for directly driving the lower boundary of global coronal models of CME events. Global operational models require continuous observations of the **vector magnetic field on the whole 4π surface of the Sun.** Future space missions need to focus on **multi-spacecraft constellations mapping out a larger area of the heliosphere.**

These developments are critical in order to enable **statistical flare and CME forecasting** (e.g. eruption probability and timing, estimation of strength and CME details, such as speed and magnetic field orientation) similar to weather prediction on Earth.